\documentclass{vgtc}                          

\ifpdf
  \pdfoutput=1\relax                   
  \usepackage{graphicx}                
  \DeclareGraphicsExtensions{.pdf,.png,.jpg,.jpeg} 
\else
  \usepackage{graphicx}                
  \DeclareGraphicsExtensions{.eps}     
\fi%

\graphicspath{{figures/}{pictures/}{images/}{./}} 

\usepackage{microtype}                 
\PassOptionsToPackage{warn}{textcomp}  
\usepackage{textcomp}                  
\usepackage{mathptmx}                  
\usepackage{times}                     
\usepackage{cite}                      
\usepackage{tabu}                      
\usepackage{booktabs}                  
\usepackage{enumitem}
\usepackage{url}
\usepackage[dvipsnames]{xcolor}
\usepackage{xspace}
\usepackage{lipsum}
\usepackage{float}
\usepackage{textcomp}
\usepackage{amsmath}
\usepackage{todonotes}
\usepackage{anyfontsize}
\usepackage{comment}
\usepackage{wrapfig}

\definecolor{gray}{rgb}{0.7,0.7,0.7}
\definecolor{orange}{rgb}{1.0,0.5,0.0}

\onlineid{0}

\vgtccategory{Research}

\vgtcinsertpkg

\title{Graph Models for Biological Pathway Visualization: \\ State of the Art and Future Challenges}

\author{Hsiang-Yun Wu\thanks{e-mail: hsiang.yun.wu@acm.org}\\ %
        \scriptsize TU Wien, Austria %
\and Martin N{\"o}llenburg\thanks{e-mail: noellenburg@ac.tuwien.ac.at}\\ %
     \scriptsize TU Wien, Austria %
\and Ivan Viola\thanks{e-mail: ivan.viola@kaust.edu.sa}\\ %
     \parbox{2.0in}{\scriptsize \centering King Abdullah University of Science and Technology (KAUST), Saudi Arabia}}

\abstract{ %
The concept of multilayer networks has become recently integrated into complex systems modeling since it encapsulates a very general concept of complex relationships. Biological pathways are an example of complex real-world networks, where vertices represent biological entities, and edges indicate the underlying connectivity. For this reason, using multilayer networks to model biological knowledge allows us to formally cover essential properties and theories in the field, which also raises challenges in visualization. This is because, in the early days of pathway visualization research, only restricted types of graphs, such as simple graphs, clustered graphs, and others were adopted. In this paper, we revisit a heterogeneous definition of biological networks and aim to provide an overview to see the gaps between data modeling and visual representation. The contribution will, therefore, lie in providing guidelines and challenges of using multilayer networks as a unified data structure for the biological pathway visualization.
} 


\begin{document}


\firstsection{Introduction} \label{sec:intro}

\maketitle

Over the past decade, several new scenarios from sciences or everyday life have benefited from formulating a relationship between entities as a graph.
For example, social networks~\cite{Moreno:1953:WSS}, which describe person-to-person relationships, are developed since 20th century.
The scientific discipline of complexity science is concerned with studying huge sets of relationships among entities, to make sense of the complex interrelationships within a particular structure or phenomenon.
Due to the rapid development of data measurements, simple graph representation is no longer sufficient to capture a real-world complex system, because the number of nodes grows prohibitively large, as well as the density of relationships in the system. Often, fortunately, the phenomenon can be structured along particular domain semantics that organize the graph into certain higher level structures. In social networks, for example, these could relate to geographical location, place of study, work, or hobbies. Such additional structure, gives rise to the concept of a multilayer network~\cite{Kivela:2014:JCN} to generalize the modeling of a complex relationship through a series of layers.

In addition to social networks, another well-known application for complexity science is analysis of network structures in biology.
The life functions are prominently organized in a relationship of interacting elements and chemical compounds, forming super complex networks of reactions carried out all over the entire life form.
These networks are large and heavily interconnected.
Biological pathways, which form a graph containing dozens of chemical elements representing a particular element of life, are the simplest relational entity of such networks.
To semantically manipulate such a network of interconnected pathways, one can segment it into multiple sub-networks, such as \emph{pathways} arranged with different functionalities.
Semantics in pathways can be understood in several ways.
One aspect could be the functionalities among the data, and another aspect could be species associated with the data.
According to the definition by Kivel{\"a} et al.~\cite{Kivela:2014:JCN},
the aforementioned aspects can be formulated as groups of layers of different types.

Unfortunately, the full set of pathways is too large to be easily handled in high quality with common graph visualization techniques.
Several research studies in visualization tend to solve the problem, by organizing aspects into clusters, such as \emph{subsystems} in biological ontology to reflect the underlying structures in the datasets.
The definition of the multilayer network facilitates us to revisit the modeling and visualization of biological pathways in a whole.
In this paper, we investigate the underlying graph modeling and visualization of conventional biological pathways,
and further research their common and different properties due to the heterogeneity in the state of the art.
Our contribution is to provide a summary to show the similar and dissimilar properties of the underlying graph models used in both pathway modeling and visualization.
This is done by a study on the collected literature and the classification through types of underlying graph data structures not only from the well-known pathway databases but also the pathway visualization methodologies.

The remainder of this paper is structured as follows:
In Section~\ref{sec:model}, we define the underlying graph structures often used in modeling biological pathways. 
Next, we investigate the visualization of biological pathway diagrams by researching hand-crafted diagrams and machine-computed layouts in Section~\ref{sec:visualization}.
In Section~\ref{sec:discussion}, we discuss and summarize some future research directions and challenges of pathway visualization, followed by a conclusion in Section~\ref{sec:conclude}.

\section{Underlying Structures of Biological Pathways} \label{sec:model}

\setlength{\intextsep}{2pt}%
\setlength{\columnsep}{8pt}%

A complex phenomenon is often formulated using a \emph{graph} to manipulate a relationship mathematically.

\noindent \textbf{Definition:} A simplest graph model can be described as a tuple $G=(V,E)$, consisting of a set of vertices $V=\{v_1, v_2, ..., v_n\}$ representing individual entities, and the mutual connectivity is represented by the edges $E=\{e_1, e_2, ..., e_m\} \subseteq V \times V$.
This is a common definition that allows us to describe the simplest relationship between entities.

The term used for describing more nested relationships between entities is often called multiplex~\cite{Battiston:2017:EPJST}, while other terminologies that describe similar concept are also used in conventional research.
Also reported by Kivela et al.~\cite{Kivela:2014:JCN} originally, those terms include \emph{multiplex}, \emph{multilevel}, \emph{multivariate}~\cite{Nobre:2019:STARs}, \emph{multidimensional}, \emph{multirelational}, \emph{network of networks} can be reframed and encapsulated by the definition of \emph{multilayer networks}.
In other words, one can use multilayer networks as an umbrella term to cover conventional formulation, nonetheless to say in the field of biological networks~\cite{Halu:2019:npj}.

This concept is not fully captured in  the early research of  biological pathways,
in which researchers tend to formulate the network as straightforward as possible based on their needs.
According to our investigation, the biological relationship is often formulated as \emph{Substrate Graph}, \emph{Bipartite Graph}, \emph{Hypergraph}, \emph{Reaction Graph}, or \emph{Stoichiometric Matrix} in biological pathways.
In this section, we summarize different graph structures that are frequently used in modelling biological pathways, and in comparison to the formulation of multilayer networks.
This will give readers an overview of the models and further understand the power of using multilayer network in the future.
We have described the simplest graph definition and will explain how it is expended to model biological pathways.

\subsection{Substrate Graph} \label{ssec:substrate}

\begin{wrapfigure}{R}{0.15\textwidth}
    \includegraphics[width=0.15\textwidth]{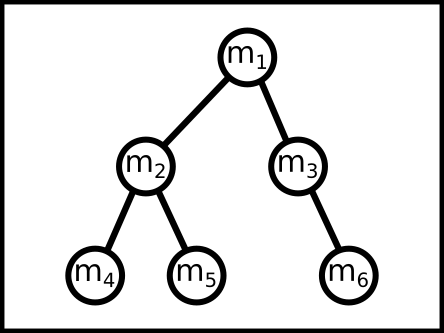}
\end{wrapfigure}

The development of computational tools in systems biology allows us to analyze the collected datasets through the power of computers~\cite{Wang:2017:SSB}.
The underlying data structure is important in the sense that it requires efficient access to the data so that the algorithms can perform efficiently.
The substrate graph is one of the pioneering graph structures.

\noindent \textbf{Definition:} A substrate graph is essentially equal to the simple graph introduced previously.
Nonetheless, since enzymes binding with chemical reactants are called substrates ($m_i$ in the figure),
for each $v \in V$ in a substrate graph represent multiple reactants together with enzymes as a single vertex.
For each $e \in E$ in a substrate graph describes a reaction between the substrates.

\subsection{K-partite Graph} \label{ssec:kpartite}

\begin{wrapfigure}{R}{0.15\textwidth}
    \includegraphics[width=0.15\textwidth]{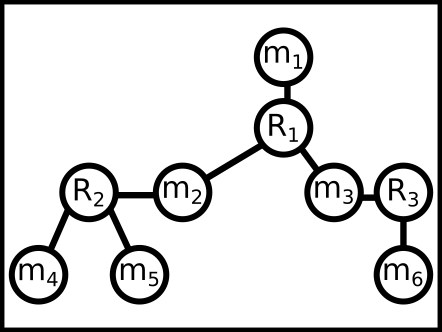}
\end{wrapfigure}

A K-partite graph is a graph whose vertices can be partitioned into $K$ different disjoint sets.
Bipartite graphs are specific types of K-partite graphs, where $K=2$,
and are common representation for biological pathways.
By definition, each vertex in a bipartite graph can be either categorized as a metabolite vertex ($m_i$) or a reaction vertex ($R_i$), but not both.

\noindent \textbf{Definition:} A graph $G$ is bipartite if and only if there exists a vertex partition, where for each $v \in V$, $V = P_1 \cup P_2$ and $P_1 \cap P_2 = \emptyset$.
Each edge $e=(v_i,v_j) \in E$, $v_i \in P_1$ and $v_j \in P_2$ so that it is guaranteed that the end vertices of an edge do not belong to the same vertex set.

\subsection{Hypergraph} \label{ssec:hyper}

\begin{wrapfigure}{R}{0.15\textwidth}
    \includegraphics[width=0.15\textwidth]{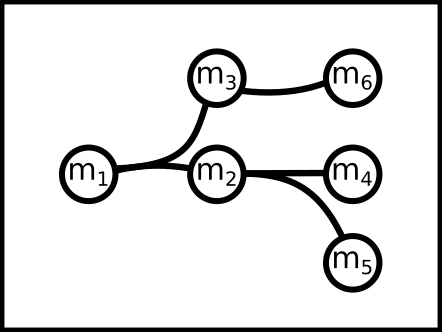}
\end{wrapfigure}

In principle, a hypergraph is a more intuitive and direct representation of biological pathways.
A hyperedge in a hypergraph can refer to a single reaction, in which participating metabolites are involved.
Although hypergraphs can be converted into bipartite graphs, and vice-versa, due to the difficulty of management, only a few tools support hypergraph representations~\cite{Wang:2017:SSB}.

\noindent \textbf{Definition:} A hypergraph $G=(V,E)$ includes a set of vertices, while $E$ is a set of non-empty subsets of $V$, namely hyperedges.
Formally, $E$ is a subset of $P$, where $P$ is the power set of $V$.

The advantage of hypergraph representation is that users can immediately understand if enzymes are involved in this reaction.
While with the substrate graph representation, it is hard to discriminate if the metabolite is a metabolite or a specific type of enzymes accelerating the corresponding reaction.

\subsection{Reaction Graph} \label{ssec:reaction}

\begin{wrapfigure}{R}{0.15\textwidth}
    \includegraphics[width=0.15\textwidth]{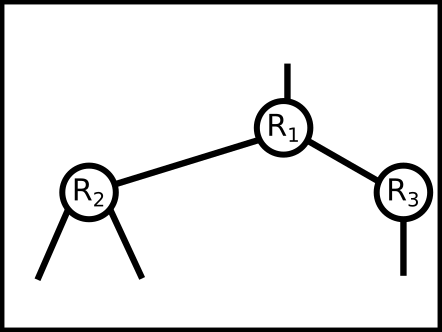}
\end{wrapfigure}

A reaction graph is also a simple graph structurally, but contains different semantics in comparison to a substrate graph.
Vertices here are distinct reactions, and the edges are metabolites involved in the network.

\noindent \textbf{Definition:} Vertices $v \in V$ in a reaction graph $G$ represent reactions in biological networks, edges $e \in E$ stand for metabolites involved.

A reaction graph is predominantly used for topological analysis, such as shortest path analysis or centrality analysis, so that users can rank graphs according to important concepts in the field~\cite{Hartmann:2015:FBB}.

\subsection{Clustered Graph} \label{ssec:cluster}

\begin{wrapfigure}{R}{0.15\textwidth}
    \includegraphics[width=0.15\textwidth]{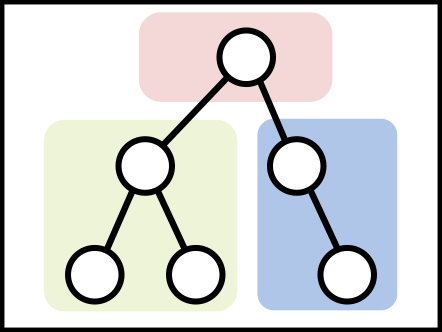}
\end{wrapfigure}

As explained by Kivel{\"a} et al.~\cite{Kivela:2014:JCN}, one can consider multilayer networks as a superset for complex networks.
\emph{Clustered graphs} is one of the interesting subsets.

\noindent \textbf{Definition:} A clustered graph is a simple graph $G$ with additional grouping information.
Each $v \in V$ in a clustered graph belong to one or more clusters $c \in C=\{c_1, c_2, ..., c_k\}$.
In other words, clusters can also form a hierarchy using a cluster tree.

In some definitions of clustered graphs, all clusters in $C$ are disjoint and form a partition of the vertex set $V$~\cite{Frishman:2004:IV}.
Nonetheless, in the context of biological pathways, clusters $c$ are not necessarily disjoint.
For example, ATP, a universal energy molecule occurring in mitochondria and cytoplasm, is often used to drive several biological reactions.
If we consider these compartments as clusters, they are overlapping since ATP can be transported from the mitochondria to the cytoplasm.
Another example would be the relationships of ATP in the biological ontology.
Since ATP occurs in many categories of biochemical reactions, including \emph{Citric Acid Cycle} and \emph{Urea Cycle}, its representation should be covered by multiple clusters in the model.
In some cases, to simplify the visual complexity of clustered graphs,
biologists duplicate unimportant vertices (e.g., vertices with high degree) to create a specific type of clustered graph, where aliases of an identical vertex only belong to a corresponding cluster.
In other words, clusters become disjoint in this case.

\subsection{Multilayer Graph} \label{ssec:multilayer}

\begin{wrapfigure}{R}{0.15\textwidth}
    \includegraphics[width=0.15\textwidth]{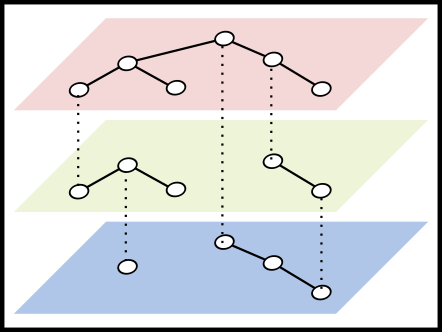}
\end{wrapfigure}

A \emph{multilayer graph} is a simple graph $G$ with additional layer information to describe real-world properties of the network in a whole.
Layers in multilayer networks are used to describe the corresponding relationships,
where each of which records the property of the corresponding relationships.
In this paper, we follow the formal definition by McGee et al.~\cite{McGee:2019:STARs}.

\noindent \textbf{Definition:}
Since each $v \in V$ can belong to several layers,
we can consider vertices as pairs $V_M \subseteq V \times L$, where $L$ is the set of associated layers.
Edges $E_M \subseteq V_M \times V_M$ indicate the connectivity of pairs $(v_i,l_p)$, $(v_j,l_q)$.
An edge is considered as an intra-layer edge when $l_p=l_q$ or an inter-layer edge when $l_p \neq l_q$, respectively.
In biological networks, we would have $L = \{l_1, l_2, l_3, ..., l_p\}$,
where $l_1$ could be metabolites occurring in mitochondria and $l_2$ could be metabolites existing in cytoplasm, and so on.
Note that some metabolites, such as $H_2O$, which occurs in both mitochondria and cytoplasm, can be connected using an interlayer edge.
This formulation becomes powerful in the sense that it covers existing  concepts and can be further used as an intermediate form to transform one concept to another, not only model wise but also visually~\cite{Kivela:2014:JCN,Gosak:2018:PLR}.

In practice, we can use multilayer graphs as a unified graph structure because the graphs described in Sections~\ref{ssec:substrate}-\ref{ssec:cluster} are specific subsets of multilayer graphs.
Thus, researchers can always convert the aforementioned graphs to multilayer graphs and again convert to the target graph data structures.
This scheme allows us to perform a systematically consistent conversion to different graph representations, as well as using the multilayer graph as a standard diagram when comparing to other visual representation.

\subsection{Stoichiometric Matrix (S Matrix)}

\begin{wrapfigure}{R}{0.15\textwidth}
    \includegraphics[width=0.15\textwidth]{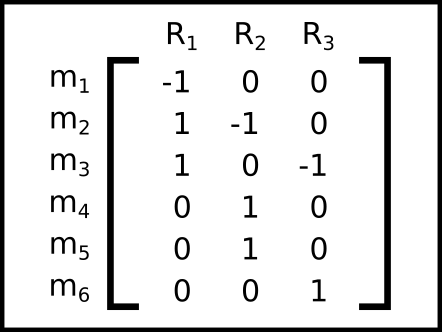}
\end{wrapfigure}

Last but not least, in addition to using graphs, the aforementioned relationship can be modelled using a stoichiometric matrix, namely \emph{S Matrix}.
With this formulation, researchers can conduct some stoichiometry-based pathway search methods which operate on the S matrix.

\noindent \textbf{Definition:} In a stoichiometric matrix, each data element (row) is a metabolite and each dimension (column) corresponds to biological reactions.
The value can be positive or negative, which indicate the stoichiometric consumption of reactants and products, respectively.
Unlike a directed graph representation that inherits the reversibility information of a reaction, this concept needs to be defined along with additional constraints, for the Stoichiometric Matrix representations~\cite{Wang:2017:SSB}.
Graphs and Matrices are sibling representations of relationship models.
Each of which has advantages and disadvantages, and can be inter-converted as well.

\section{Biological Pathway Visualization} \label{sec:visualization}

The purpose of developing pathway visualization is to convey the underlying knowledge of the data models to the users effectively.
Ideally, the visualization is expected to reflect this information intuitively.
Unfortunately, pathway visualization is still a challenging problem due to the size and the complexity of biological models.
Moreover, the heterogeneity of biological networks makes the problem even worse, since it becomes difficult to develop a unified framework that supports various types of information.
In this section, we aim to summarize the existing visualization techniques in comparison to the models described in the previous section.
The focus includes both the overview diagram from the public biological databases and the machine-generated pathway diagrams.

\subsection{Diagrams from Biological Databases}

Biological databases enable researchers to store, analyze, and share biological information systematically~\cite{Wren:2008:BI,Rigden:2017:NAR}.
Metabolic pathway and protein function databases often give an overview map~\cite{Noronha:2017:bioinfo} to firstly demonstrate the high-level structures of the organized biological information.
\emph{BioCyc Database Collection} is a Pathway/Genome Database, which collects metabolic and genome pathways of organisms.
Its overview diagram, \emph{Pathway Collage}~\cite{Paley:2016:MBCBI}, contains user-specified pathways for an organism and can be arranged and customized on their user interfaces.
This is a semi-automatic approach~\cite{Karp:2015:BiB}, where
smaller graphs are automatically computed~\cite{Karp:1994:LISP}, followed by a manual arrangement to avoid overlaps~\cite{Paley:2016:MBCBI}.
Vertices here are small-molecule metabolites or proteins, and edges biochemical or transport reactions.
The underlying graph here is a multi-faceted clustered graph.

\emph{BRENDA}~\cite{brenda} is an information system, which allows the users to search, filter, and retrieve enzyme functions and where a molecule is involved in the enzyme functions.
Since 2017, the team introduced an overview map of the metabolic pathways, to visualize the enzyme and ligand information.
The map has been created manually drawn using the network editor Cytoscape~\cite{Shannon:2003:cytoscape},
in order to create a map which is familiar to biochemists and similar to the maps in biological textbooks.
The visualization is in a top-down fashion, where they show the category (approx. seven categories) using color boxes, and expanding the detailed pathways when clicking on the boxes.
The current map contains ligand and enzyme as vertices and edges as reactions; therefore, it is a combination of clustered graph and bipartite graph representations.

\emph{HMDB}~\cite{Wishart:2009:NAR} contains detailed information about small molecule metabolites found in the human body.
The visualization suite, SMPDB~\cite{Frolkis:2009:NAR}, provides an interactive interface to small human metabolic pathways, but the team did not integrate them into an overview map.
The pathways in the database are hand-drawn pathways specific to humans.
A vertex can be a metabolite or enzyme, and edges are reactions in the biological processes.
The difference here is that the designer added background icons such as mitochondria or endoplasmic reticulum, to describe the spatial location of the reactions in the physical world.
The visualization uses a bipartite graph representation with annotated image labels.

\emph{KEGG PATHWAY Database}~\cite{Kanehisa:2000:nar} collects manual-drawn pathway maps, which represent our knowledge on the molecular reactions.
The vertices in the overview map are metabolites and edges are reactions, while the involved enzymes are integrated in a singe edge representation.
The diagram is clean with only category information.
The users need to click on a vertex or an edge to explore further details due to the complexity of the diagram.
This overview diagram is a combination of clustered graph representation together with a user interface.
\emph{MANET database}~\cite{Kim:2006:BMCBI} is a database that allows users to trace the evolution of protein architecture in biomolecular networks.
The interface supports scripts, query, and statistical analysis.
 The researchers utilize the pathway maps created from the KEGG database~\cite{Kanehisa:2000:nar},
and thus do not have their own specification.

\emph{Reactome}~\cite{Konstantinos:2017:bio} is the pioneering database, which incorporates fully automatic layout algorithms to assist navigation of human metabolic pathways.
The system first provides a hierarchical visualization to show the relationship of biological pathway categories.
Once the user selects a category, the system will zoom in and show a detailed pathway diagram.
The collection of pathways is organized to show the relationship of pathway categories, then the detailed pathways, unlike the other databases, which tend to create a big overview map based on species.
Vertices represent the participants of reactions, which are categorized into ten types, including complex, protein, and RNA.
Edges in the pathway diagrams are reactions.
The database uses a hypergraph representation for the information.

\emph{ReconMap}~\cite{Noronha:2017:bioinfo} is developed based on the genome-scale reconstruction of human metabolism, and provides an interface to access its content and associated omics data and simulation results.
It is known as the largest hand-crafted human metabolic pathway map that facilitates an overview of the major biological processes in the human body.
The vertices in the diagram are metabolites which are categorized based on their semantics,
and edges represent biological processes.
This overview diagram is arranged in a way that reactions happen in the same compartment in a cell are strictly placed within a certain representative region.
To achieve this, some high-degree vertices are duplicated.
This diagram is a combination of a clustered graph with disjoint clusters and a hypergraph representation.

\emph{WikiPathways}~\cite{Slenter:2017:NAR} is another collection of knowledge concerning biological pathways.
Similar to HMDB, WikiPathways collects many promising pathway information but does not visually integrate them.
Vertices here are metabolites and edges are reactions.
The diagrams are using hypergraph representation.

Other well-known pathway diagrams (which are not databases) such as \emph{Roche Biochemical Pathways}~\cite{roche}, even embed chemical structures to guarantee the high quality of the pathway map.
The diagram is considered as a hypergraph representation with large number of vertex annotations.

\subsection{Software-Assisted Pathway Diagrams}

Since new biological pathways are unceasingly introduced and added to pathway databases,
pathway visualization has developed a series of variations to support pathway analysis.
Recently, Murray et al.~\cite{Murray:2017:mbc} proposed a visualization task taxonomy for the analysis of biological pathway data.
They summarize that relationship tasks are considered as the most common and essential tasks in their investigation,
while they also point out that existing network visualization tools are not that fully suitable for biological pathways due to their limited capability~\cite{Vehlow:2015:BMC,Paduano:2016:VINCI}.
The more constraints added to the layout, the higher running time is needed.
Since the layout problem is still resource-consuming, most of the techniques can handle graphs with up to $500$ vertices.

For this reason, pathway visualization is commonly categorized into two types, which are an interactive exploration scheme on a static map, and a fully automatic layout approach.
For the first case, the static map must be created in advance.
Scientists need to either use pathway editors, such as CellDesigner~\cite{Funahashi:2003:celldesigner}, SBGN-ED~\cite{Klukas:2010:BI}, or Newt~\cite{Sari:2015:PLOS} to create the pathway diagrams, or adopt conventional network analysis tools, such as Gephi~\cite{Bastian:2009:AAAI} and Cytoscape~\cite{Shannon:2003:cytoscape} to apply standard layout approaches.
While the maps are often of better visual quality, creating static maps often requires a lengthy trial and error process,  until the design is finally satisfactory.
With a nice hand-crafted map, biologists can rely on powerful user interactions with the maps to support a comprehensive understanding of the datasets.
For example, basic interactions such as semantic zooming~\cite{Gawron:2016:minerva} and visual aggregation~\cite{Dogrusoz:2006:CG, Dogrusoz:2018:PLOS} have been investigated to analyze large networks.
Note that interaction has been proved useful for large network visualization, but it may also generate extra overhead for doing simple connectivity tasks~\cite{Yoghourdjian:2018:CoRR}.
Nevertheless, interaction provides us an alternative to expand and collapse the visualization in order to navigate and access our target of interests.

On the other hand, the second type requires sophisticated layout algorithms by introducing several pathway-specific constraints.
Several research works focus on the visually pleasing and well readable layout of small biological networks.
Gerasch et al. rebuild small KEGG pathway maps~\cite{Gerasch:2014:pvis} by removing unnecessary elements in the diagram, to conserve the original KEGG layout.
The approach is built on top of hierarchical sub-networks to support user interaction.
Jenson and Papi present MetDraw, an approach that generates pathway diagrams using a familiar graph library GraphViz, together with overlaying omics data~\cite{Jensen:2014:bioinformatics}.
Li and Kurata formulate the pathway diagram using an annealing procedure~\cite{Li:2005:bi}.

The idea is to arrange vertices on a 2D square grid to exhibit cluster information and produce a compact diagram.
Bachmaier et al.~\cite{Bachmaier:2014} also summarize several pathway layout approaches extended from the well-known force-directed technique.
Wu et al.~\cite{Wu:2019:BMCBI} introduced a new design for biological pathways, where an urban map metaphor is used to maintain the readability of low-level and high-level relationship for larger networks, such as human metabolic pathways.
Several pathway visualization toolkits are also investigated in the state-of-the-art literature~\cite{Suderman:2007:bio,Pavlopoulos:2008:BDM,PGehlenborg:2010:NM,Bachmaier:2014}.
Most automatic layout approaches rely on constrained layout techniques, which often assume the input graph is simple and thus restricts the visual representation of biological pathways.
Since the multilayer network concept is considered as a framework for the graph formulation of a complex system, it thus can serve as an intermediate structure to convert between different graph models and visualizations.

\section{Future Challenges of Pathway Visualization} \label{sec:discussion}

After investigating the commonly used pathway databases, we summarize the challenges as follows:

\begin{itemize}[leftmargin=0pt]
\setlength\itemsep{-1mm}

    \item[] \textbf{Structural Motifs}:
    Structural network motifs are simplified and representative patterns that show the most common structures in the biological networks. Besides, the motifs that abstract the shapes of metabolites can provide insight during simulation and analysis~\cite{roche}.
    One interesting aspect of these motifs is that they provide a visual summary for the structures so that the visualization is clean and readers can intuitively understand the knowledge behind it~\cite{Viola:2017:TVCG}. 
    Technically, this introduces another challenge since the screen space reserved for the motifs should be balanced with the network layout, which implies a difficult layout challenge for multilayer networks.

    \item[] \textbf{Global vs. Local Relationship for Biological Tasks}:
    As reported by Murray et al.~\cite{Murray:2017:mbc}, detailed relationship information, as well as the corresponding aspects, are expected to be clearly presented for analysis or educational purposes in a pathway visualization.
    This implies that any well-organized layer structures should be compatible with the high-level and detailed-level connectivity of biological pathways.
    Aside from pathway modeling, a visual hierarchy in multilayer networks should be explicitly developed in order to maximally reduce the cognitive work load of the visualization users.

    \item[] \textbf{Scalability and Interactivity}:
    Scalability and interactivity are key factors for all of the visualization techniques.
    This is especially significant for pathway analysis because the new findings impact the field frequently.
    A typical example is the study of \emph{glucose}, which was considered as a fast supply of energy initially but later on is also found to have a strong activation of cancer metabolism.
    This changes the topology of the graph along time.
    We, therefore, have a strong need for dynamic layout approaches, and this must be applied to large graphs since there are thousands of metabolites and reactions in human metabolism.

    \item[] \textbf{Application-Oriented Approach}:
    In the past decades, scientists often appreciate more general visualization techniques for networks so that they can be easily applied to different research fields.
    Nonetheless, since the field in biology is becoming more and more complex, and several properties in the domain may not exist in other domains, a domain-specific approach is required.
    This is because a general approach is not specialized enough to support the regular analysis, which also generates a lot of overhead for domain scientists.

    \item[] \textbf{Dynamics and Time-dependency}:
    Time is another important factor.
    How to incorporate time as an axis in multilayer networks is still an open research question because it adds another complexity in both modeling and visualization.

    \item[] \textbf{Uncertainty}:
    The current biological network is deterministic based on experiment settings.
    However, in the real-world case, some pathway interactions can be significant, while some are not.
    This information could provide more insight into the effectiveness of drugs during development and should be considered as one primary direction in the next research generation.

    \item[] \textbf{Comparative vis, translation between layout and visualization}:
    Since the data collected in biology is heterogeneous, researchers would like to compare similar and dissimilar pathways across different organisms, cell compartments, and species, to investigate potential solutions for the interesting research questions.
    Multilayer networks facilitate this opportunity since they integrate the heterogeneity into a standard scheme, and therefore enable researchers to retrieve and provide a framework for the application of advanced visualization techniques.

\end{itemize}

\section{Conclusion and Future Work} \label{sec:conclude}

In this paper, we summarize the roles and challenges of various graph structures used in biological pathway modeling and visualization.
Meanwhile, we collect frequently used hand-crafted and machine-generated pathway maps, to give an overview on how the graph, as well as multilayer networks,  can be employed in the domain. 
Finally, we summarize the potential research challenges and directions for multilayer network visualization, in the context of biological pathways.

In the future, we attempt to investigate, step-by-step, the aforementioned challenges, to further develop a unified biological pathway visualization technology.
Our first task is to conduct a general framework that integrates various graph models by using multilayer networks as central data structures.
This can be an exciting research direction in the sense that scientists can explicitly select and switch the corresponding graph visualization based on their preferences.
This also facilitates the possibility to convert or translate among different network visualizations to cognitively support data analysis for various purposes.

\acknowledgments{
The project has received funding from the EU Horizon 2020 research and innovation programme under the MSCA grant No. 747985, from the Vienna Science and Technology Fund (WWTF) grant No. VRG11-010, from the Austrian Science Fund (FWF) grant No. P31119, and from King Abdullah University of Science and Technology (KAUST) through award BAS/1/1680-01-01.
}

\clearpage

\bibliographystyle{abbrv-doi}

\bibliography{paper}
\end{document}